\documentclass[preprint,apjl]{emulateapj}

\newlength{\starprofwidth}
\newlength{\profwidth}
\newlength{\agnwidth}
\newlength{\profwidthtwo}
\newlength{\agnwidthtwo}
\newlength{\colwidth}
\newlength{\fullwidth}
\newlength{\twothirdwidth}
\newcommand{\galfit}{{\sc galfit}}
\newcommand{\ferengi}{{\sc ferengi}}
\newcommand{\kcorrect}{{\tt k\_correct}}

\newcommand{\Ngal}{96}
\newcommand{\sersic}{S\'ersic}

\shorttitle{FERENGI: Redshifting galaxies}
\shortauthors{M.~Barden, K.~Jahnke \& B.~H\"au\ss ler}

\begin{document}

 \title{FERENGI: Redshifting galaxies from SDSS to GEMS, STAGES and COSMOS}

 \slugcomment{Accepted September 26, 2007 to ApJL}

\author{M.~Barden\altaffilmark{1,2}, K.~Jahnke\altaffilmark{1}, 
        B.~H\"au\ss ler\altaffilmark{1}}

\altaffiltext{1}{Max-Planck-Institut f\"ur Astronomie, K\"onigstuhl
17, D-69117 Heidelberg, Germany}
\altaffiltext{2}{Institute of Astro- and Particle Physics, University of
Innsbruck, Technikerstra\ss e 25, A-6020 Innsbruck, Austria} 
\email{marco.barden@uibk.ac.at, jahnke@mpia.de, boris@mpia.de}

\begin{abstract}
We describe the creation of a set of artificially ``redshifted'' galaxies in 
the range $0.1<z<1.1$ using a set of $\sim$100 SDSS low redshift ($v<7000$ 
km\,s$^{-1}$) images as input. The intention is to generate a training set of 
realistic images of galaxies of diverse morphologies and a large range of 
redshifts for the GEMS and COSMOS galaxy evolution projects. This training 
set allows other studies to investigate and quantify the effects of 
cosmological redshift on the determination of galaxy morphologies, 
distortions and other galaxy properties that are potentially sensitive to 
resolution, surface brightness and bandpass issues. We use galaxy images from 
the SDSS in the $u$, $g$, $r$, $i$, $z$ filter bands as input, and computed 
new galaxy images from these data, resembling the same galaxies as located at 
redshifts $0.1<z<1.1$ and viewed with the Hubble Space Telescope Advanced 
Camera for Surveys (HST ACS). In this process we take into account angular 
size change, cosmological surface brightness dimming, and spectral change. 
The latter is achieved by interpolating a spectral energy distribution that 
is fit to the input images on a pixel-to-pixel basis. The output images are 
created for the specific HST ACS point spread function and the filters used 
for GEMS (F606W and F850LP) and COSMOS (F814W). All images are binned onto 
the desired pixel grids (0\farcs03 for GEMS and 0\farcs05 for COSMOS) and 
corrected to an appropriate point spread function. Noise is added 
corresponding to the data quality of the two projects and the images are 
added onto empty sky pieces of real data images. We make these datasets 
available from our website, as well as the code -- \ferengi: ``Full and 
Efficient Redshifting of Ensembles of Nearby Galaxy Images'' -- to produce 
datasets for other redshifts and/or instruments.
\end{abstract}


\section{Introduction}\label{sec:intro}
In the current era of observational astronomy the size of galaxy datasets
means that number statistics starts to lose its spot as number one source of
uncertainty in galaxy evolution studies. With the availability of large wide
field galaxy surveys as the Sloan Digital Sky Survey \citep[SDSS][]{york00} or
2dF Galaxy Redshift Survey, and the deep, space based high resolution projects
like the Hubble Ultra Deep Field, GOODS, GEMS and STAGES projects and the 
Cosmic Evolution Survey (COSMOS) with their $10^4$ to $10^6$ galaxies, other 
error sources become vital to understand.

If we want to understand the buildup of galaxies with their intricately linked 
evolution in stellar and black hole mass, luminosities, colors, morphological 
types, and the alternations between interaction and relaxation, we have to 
understand what the tools we apply really measure. Any given galaxy will look 
different when viewed with different instruments or when located at different 
redshifts, due to cosmological dimming and changes in rest-frame bandpass. 

This makes the qualitative or quantitative classification of e.g.\ galaxy 
morphology non-trivial to compare for different redshifts or filters. As one 
example faint disks visible at one redshift will fade from the observer's 
view at larger distances, not only affecting eyeball-classifications of 
morphology but also automatic classifier software. Moreover, low surface 
brightness signs of interaction and structures, such as bars which are not 
prominent in the rest-frame ultraviolet, will not be visible at higher 
redshfits.

If the evolution of galaxies is to be studied via merger statistics,
morphology-segregated evolution, star formation histories, and type-specific
luminosity functions, all morphological classifiers have to be calibrated to
deliver comparisons of similar quantities at all redshifts, taking into
account bandpass-dependent properties and changes in signal-to-noise ratio.

From the analysis of HST survey data we know that the average surface 
brightness of the disc galaxy population fades with time 
\citep[e.g.\ ][]{lilly1998,barden2005}. The change is approximately 1-1.5mag 
depending on rest-frame bandpass over the redshift interval $0<z<1$. To some 
degree this surface brightness evolution counters the cosmological dimming 
and helps detecting low surface brightness features. Thus, if this effect is 
not taken into account, predictions about the recoverability of structural 
parameters or classifications are overly pessimistic. 

For a few projects in the past, codes were created that would include some 
of these effects for specific applications or datasets 
\citep[e.g.][]{abraham1996a,abraham1996b,giavalisco1996,bouwens1998,
takamiya1999,burgarella2001,kuchinski2001,vdbergh2002,lisker2006}. 
However, to our knowledge no codes or datasets that include geometrical and 
cosmological bandpass shifting effects are currently publicly available. In 
the present article we describe the creation of artificial image data sets in 
the range $0.1<z<1.1$, computed from low-$z$ SDSS galaxies by artificial 
``redshifting''. Since we refrain from adding any evolutionary models but 
purely apply cosmological changes in angular size, surface brightness and 
filter bandpass, this dataset shows exactly how such galaxies will appear 
when observed from cosmological distances, at which redshift certain features 
become undetectable, and any quantitative classifier procedure can be tested 
for dependency on redshift effects.

Our code \ferengi, ``Full and Efficient Redshifting of Ensembles of Nearby 
Galaxy Images'', and the present datasets were originally created for the
morphological classification of active and inactive galaxies in the GEMS
\citep{rix04}, STAGES \citep{grayinprep}, and COSMOS \citep{scov06} projects. 
The datasets for the HST ACS image characteristics of the GEMS and COSMOS 
project are freely available from our website\footnote{Website for retrieval 
of simulated datasets and the code 
\ferengi: \url{http://www.mpia.de/FERENGI/}}. As many studies 
might benefit from such data we also make the \ferengi\ code available on the 
same webpage for others to use with different input samples, redshifts, and 
instrument characteristics.

In the following we describe the input data in Section~\ref{sec:input}, the
redshifting procedure including the basic cosmological formulae that enter,
and the bandpass shift in Section~\ref{sec:redshifting}, and the creation of
realistic images from this information in Section~\ref{sec:noise} including
example results (Section~\ref{sec:results}). Next, we present a series of 
tests characterising the robustness and accuracy of the code 
(Section~\ref{sec:tests}). In Section~\ref{sec:conclusions} we discuss 
limitations of this procedure.

All examples and cosmology-dependent numbers given here are computed
assuming a flat universe with $\Omega_\Lambda=0.7$ and
$h=H_0/(100\,\mathrm{km\,s^{-1}\,Mpc^{-1}})=0.7$.

\section{Input galaxy sample}\label{sec:input}
The basic idea is to convert images of well resolved, low-redshift galaxies to
images simulating the same galaxies at higher redshift. Input and output images
can differ in assumed redshift, pixel size, point spread function (PSF) and
noise properties. We explicitely compute two versions, with and without 
luminosity / surface brightness evolution terms. When purely considering 
instrumental and cosmological effects, users of such datasets can apply his 
or her tools to the exactly same galaxies, only located at different 
redshifts for calibration purposes. When including evolution, more realistic 
galaxies are created. For the current study we have implemented a linear 
scaling with redshift to make sources brighter at high redshift.

For this project we require input galaxy data with two main properties: i)
Sufficient sampling: The combination of distance to the galaxy, pixel size and
width of the PSF must be sufficient for the target pixel size and PSF width at
the lowest target redshift (e.g.\ $z\ge 0.1$). ii)
Information about the spectral energy distribution: The task needs homogeneous
multiband imaging used for interpolation between filters to compute fluxes in
the target filter band at the desired redshift without extrapolation.

\begin{figure}[t]
\plotone{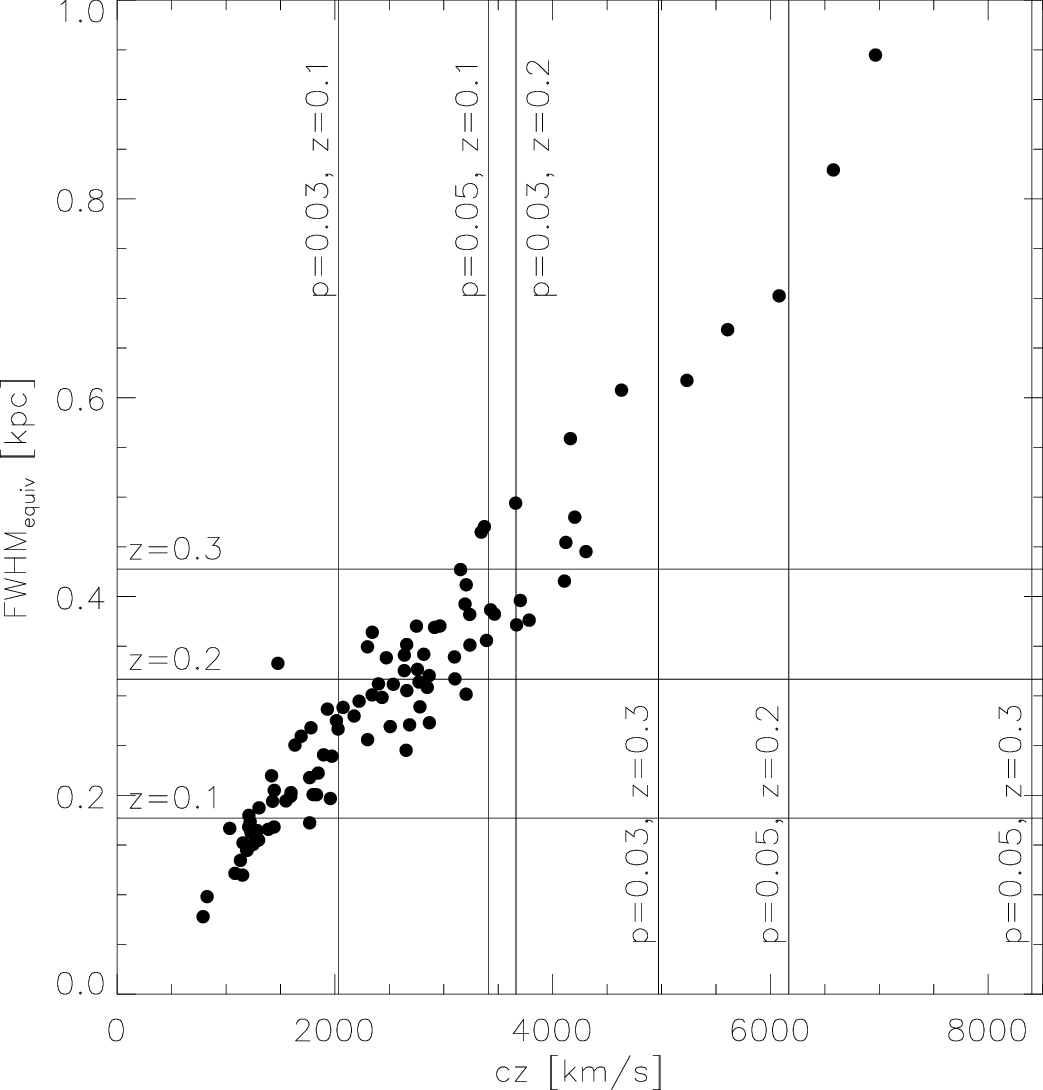}
\caption{\label{fig:sampleselection} 
Sample selection. Shown is the width of the PSF of initially selected SDSS
galaxies as a function of recession velocity. We give the width in equivalent
linear dimensions, i.e.\ the linear size of the angular FWHM-width of the PSF 
in the SDSS $r$-band, at the distance of the galaxy. The horizontal lines
correspond to upper selection limits for the input width to not exceed the ACS
PSF width in the output images, at $z=0.1, 0.2$ and $0.3$, respectively. The
vertical lines mark the $cz$ upper selection limits from input pixel size, to
not exceed the output pixel sizes of 0\farcs03 and 0\farcs05 at the given
redshifts.
}
\end{figure}

The best source for such data are nearby galaxies from the SDSS. This survey
provides imaging data in its own $u$, $g$, $r$, $i$, $z$ filters with
0\farcs396/pixel sampling. We compute the maximum distance for input galaxies
from the two requirement to i) simulate images with 0\farcs05 and 0\farcs03
pixel size at a minimum redshift of $z=0.1$, and ii) to have an output PSF not
already broader than the $\sim0\farcs096$ (FWHM) PSF of the ACS WFC drizzled 
images in the F606W or F814W filters. Figure~\ref{fig:sampleselection} shows 
the resulting limit in $cz$ and the limit from the width of the PSF for a 
sample of SDSS galaxies and their PSF conditions. Both criteria select a 
similar set of galaxies, corresponding to limiting recession velocities of 
$v\la2000, 3700$ and $5000$~km\,s$^{-1}$ for $z_\mathrm{out,min}=0.1, 0.2$ 
and $0.3$, respectively, and $0\farcs03$ output pixel size. For pixel size 
0\farcs05 the limits change to $v\la3400, 6200$ and $8400$~km\,s$^{-1}$.

These technical aspects define the boundary conditions of our sample 
selection. We emphasise that at this point we do not aim to select a sample 
that is complete or representative for the general population of galaxies in 
a statistical sense. The sample selection only intends to span a large range 
of morphological types and common features, targeting large, rather luminous 
galaxies that could still be observed at larger $z$.

With this intention we make a selection of galaxies from the SDSS survey area,
via the HyperLeda \citep{Patu89,Pru98} catalogue facility, with the following
parameters: recession velocity $200\le v_{\mathrm{vir}}\le7000$ corrected for 
Virgo infall (208 km\,s$^{-1}$), apparent isophotal diameter $D_{25}\ge1$ 
arcmin, and apparent magnitude $B_T\le16$, and added a total $B$-band 
magnitude cut at $M_{B_T} \le -19.5$.

Images were taken from the data release 4 (DR4) of the SDSS \citep{abaz05}. As
the SDSS imaging data is not targeted on individual galaxies, a fraction of
about 40\% of the selected galaxies extends beyond the borders of the
680$\times$590 arcsec$^2$ SDSS field of view. We excluded such galaxies as
input after visual inspection.

\begin{figure}[b]
\plotone{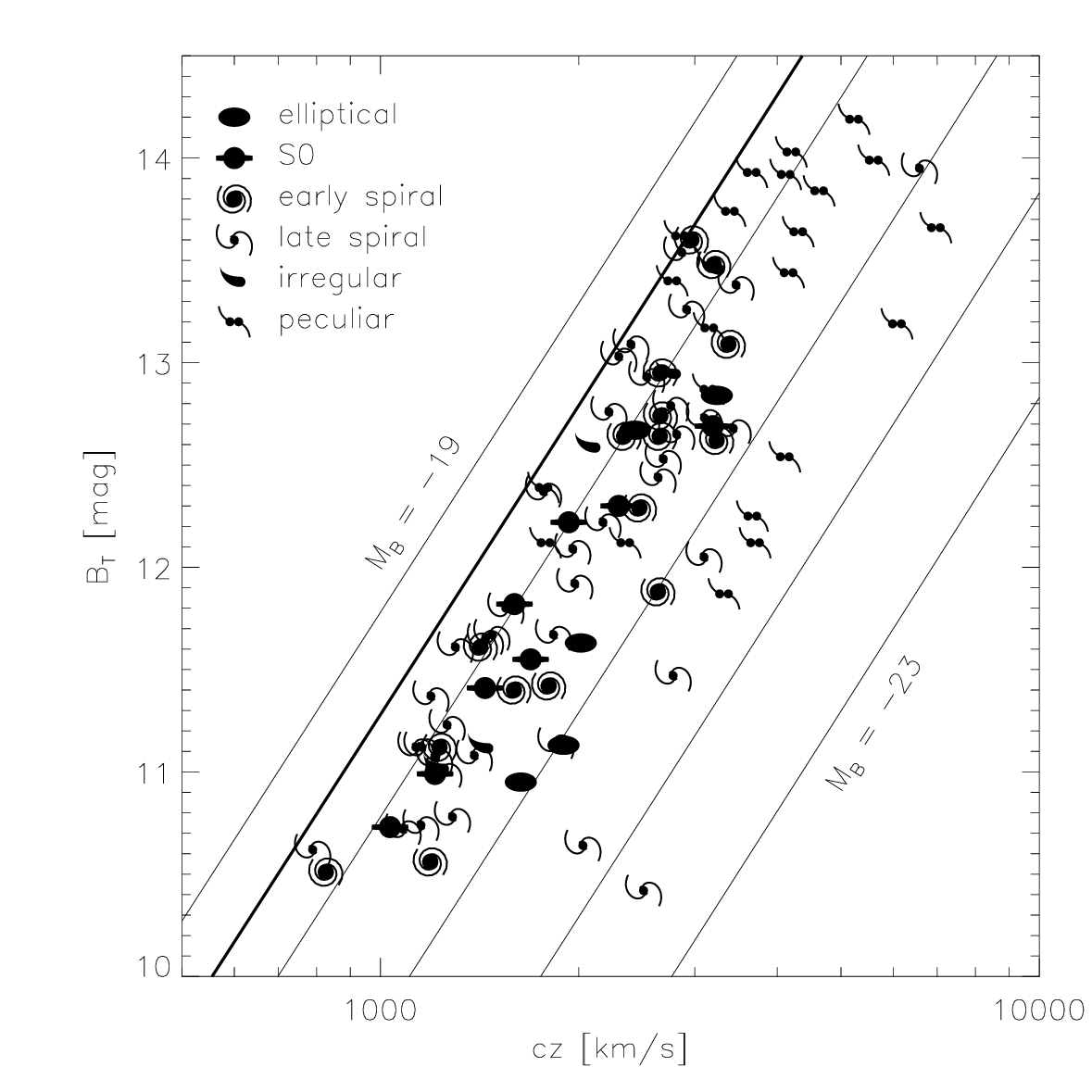}
\caption{\label{fig:sample}
Hubble diagram of the selected input sample. Indicated are the approximate
morphologies to give an overview over the diversity of the sample. The thin
diagonal lines correspond to approximate absolute magnitudes $M_B$ of $-18$ to
$-23$, spaced at 1~mag interval. The thick lines is our imposed selection cut 
at $M_B=-19.5$.
}
\end{figure}

In total we selected a sample of \Ngal\ galaxies with $cz<7000$
km~s$^{-1}$ and very heterogeneous morphologies. Figure~\ref{fig:sample}
shows the Hubble diagram of the input sample. In the same figure and in
Table~\ref{tab:sample} we give the distribution of morphological
classification according to the {\em Third Reference Catalogue of Bright
Galaxies} \citep[RC3,][]{deva91} to illustrate the sample
composition. Note, that the class ``peculiar'' was manually selected for
showing signs of ongoing mergers, or close or overlapping galaxies.

\begin{table}[t]
\begin{center}
\caption{Distribution of RC3 morphological classes.\label{tab:sample}}
\begin{tabular}{lc}
\tableline
\tableline
Class\tablenotemark{a}&Number of galaxies\\
\tableline
E & 5 \\
S0 & 8 \\
early S\tablenotemark{b} & 16 \\
late S\tablenotemark{c} & 40 \\
Irr & 5 \\
pec\tablenotemark{d} & 22 \\
\tableline
\end{tabular}
\end{center}
\tablenotetext{a}{Coded revised Hubble type according to RC3}
\tablenotetext{b}{Hubble types S0/a--Sb}
\tablenotetext{c}{Hubble types Sbc--Sm}
\tablenotetext{d}{manually classified (multiple source and/or 
merger signatures)}
\end{table}

As in the local universe few galaxies show strong signs of interactions or 
even mergers, we gradually extended our selection range from $500<cz<3000$ out 
to $cz<7000$. At larger distances we only included ``peculiar''-looking 
sources like strong interactions or mergers. However, we also included some 
unrelated objects that have small projected radii or even overlap. Our 
intention here was to demonstrate the increasing difficulty at high redshift 
to morphologically discern interaction from projection effects.

\section{Redshifting procedure}\label{sec:redshifting}
The \ferengi\ redshifting procedure has two components, cosmological angular 
size and surface brightness changes on one side and bandpass shifting on the 
other. After computation of these cosmological effects we correct for PSF 
effects and add noise.

\subsection{Angular size and surface brightness}
When converting from an input image with pixel size $p_i$ of a galaxy at
redshift $z_i$ to an output redshift $z_o$ and desired pixel size $p_o$ both
angular sizes and surface brightnesses have to be modified for distance and
cosmological effects. Using $\tan(a) \simeq a$ for small angles, the angular
size $a$ of a given linear dimension changes as
\[
 \frac{a_o}{a_i} = \frac{d_i/(1+z_i)^2}{d_o/(1+z_o)^2}
\]
with the luminosity distance $d$. Since the input image will be rebinned to the
output pixel size, angular sizes in units of pixels $n$, and thus image
rebinning factors, are
\begin{equation}
 \frac{n_o}{n_i} = \frac{d_i/(1+z_i)^2}{d_o/(1+z_o)^2}\frac{p_i}{p_o}.
\end{equation}

The flux in each pixel is subject to surface brightness dimming. If we require
the galaxy absolute magnitude to be conserved
\[
M=m_i-5\,\log(d_i)-c=m_o-5\,\log(d_o)-c
\]
%
we find a relation for the ratio of observed fluxes $f$ of the in- and output
images
\begin{equation}
2.5\,\log\left(\frac{f_o}{f_i}\right)=m_i-m_o=5\,\log\left(\frac{d_i}{d_o}
\right) \Leftrightarrow \frac{f_o}{f_i}=\left(\frac{d_i}{d_o}\right)^2,
\end{equation}
which gives the standard bolometric surface brightness dependence of
$(1+z)^{-4}$. The finite filter width is taken care of in the next component.

Note, that in order to account for the bandpass shift it is imperative to 
match object positions in the input images. This matching is a complex 
process, which potentially includes shifting, rotation and scaling of the 
input data. Moreover, the PSFs of the data should match. In the case of 
strongly varying PSFs colour terms may be introduced in the bandpass shifting. 
When combining e.g.\ GALEX and SDSS data, to extend the wavelength 
baseline to the UV, the users would have to prepare the input image 
accordingly. They have to shift all images to a common position, scale the 
GALEX image to the SDSS pixel scale, possibly rotate the GALEX image to match 
the SDSS frame and smooth all SDSS images to the GALEX FWHM. The so-processed 
images are then input to \ferengi. Thus, preparing the input appropriately, 
\ferengi\ is not limited in the combination of telescopes, filters, or 
instruments.

\subsection{Bandpass shift}
While the above rebinning and flux rescaling takes care of all geometrical
effects, bandpass shifting and stretching still has to be added. The 
cosmological redshift will (by definition) shift the observed restframe 
bandpass for a given observing filter as a function of redshift as well as 
change the size of a given filter. In addition we want to have the option to 
produce images for a range of optical observed filters.

For this task we input spectral information in the form of multiband imaging. A
combination of observers filter and redshift defines the desired rest-frame
filter curve. For each pixel of the input frame we calculate the expected flux
for this rest-frame filter, by interpolating between multiband information on
this pixel. This task is facilitated by the routine \kcorrect\
\citep[v4.14,][available for IDL and C]{blan03} that was written to fit 
spectral templates to a set of multiband images. We use \kcorrect\ to 
determine a best template for individual pixels in an image, and from this 
the flux in each pixel for a given filter. We so construct a rest-frame 
filter image. This bandpass shift and the above size-rebinning are 
interchangeable in order. To minimize the computational effort and to reduce 
noise we first recomputed the (coarser) redshifted images and then applied 
the bandpass shift. Also, only pixels with flux exceeding 2 times the rms of 
the background are input to \kcorrect\ (optional feature). The remaining 
pixels receive a flux-weighted $K$-correction computed from the bright 
pixels. While this could become more important at low S/N levels, it showed 
not to have a significant effect for the currently chosen bright sample of 
galaxies.

Note, the template set incorporated in \kcorrect\ stems from \citep{bc03} 
models. These model templates cover the rest-frame wavelength range from UV to 
IR (600\AA\ to 320$\mu$m). As long as the input filters together with the 
targeted output redshift fall within this range any combination is possible. 

\begin{figure*}[htp]
\plotone{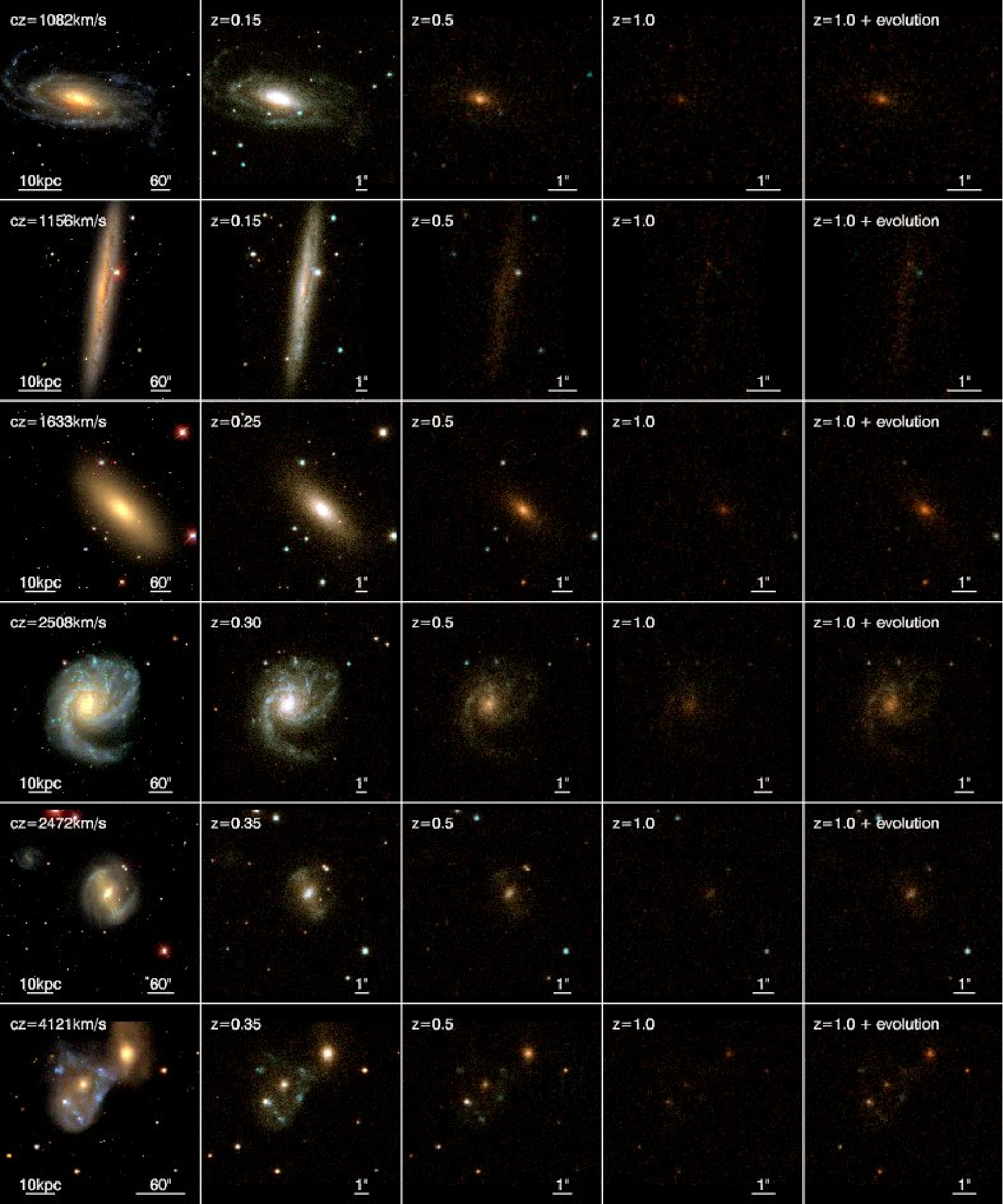}
\caption{\label{fig:examples}
Redshifting examples. From left to right the first four panels in each row show
the original SDSS $gri$-colour composite image and the redshifted GEMS 
F606W-F814W false colour images at $z=0.15-0.45, 0.5, 1.0$. The fifth column 
includes magnitude evolution by $\Delta m=-1\times z$ mag at redshift $z=1$. 
Apparent scales are indicated in each panel; physical scales are the same in 
each panel and therefore only marked in the leftmost panel.
}
\end{figure*}

\section{Point spread function and noise}\label{sec:noise}
Two essential ingredients for realistic images are i) to mimic the resolution 
of the real data by convolution with an appropriate PSF and ii) to generate
realistic noise levels for galaxy and background.

The input SDSS galaxies were observed under varying seeing conditions with an
average PSF width of 1\farcs4 FWHM corresponding to $\sim$3.5 SDSS camera
pixel. After application of the redshifting procedure the intrinsic PSF 
is of finite width and non-negligible. While for point- or composite sources as
type~I AGN the detailed knowledge about the PSF is essential, the requirements
are not so stringent for extended galaxies. However, as the simulated dataset 
is meant to calibrate quantitative morphological analysis methods and 
programs, it is important to create a PSF that is sufficiently similar to the 
real PSF for the task in question.

In light of this we attempt to reach a final PSF as close as possible in shape
and width to the ACS PSF in the desired band. This involves the creation of a
convolution kernel that will produce the ACS PSF shape from the input PSF 
shape. As the width and geometric shrinking of the PSF depends both on input 
and target redshift, and the input PSF varies, this kernel needs to be 
recomputed for each input galaxy and output redshift. This is done by a 
transformation into Fourier space, suppression of noise by Wiener filtering, 
and subsequent division of the spectra of the two PSFs. After transformation 
of this quotient back into the spatial domain we receive the function that is 
needed to convolve the input image to reach a PSF close to the ACS PSF. This 
is mathematically equivalent to a deconvolution of the output PSF with the 
desired input PSF. For the comparably high S/N PSF in these data this works 
fairly well. Note, that the filter of the input PSF is chosen from the set of 
SDSS filters minimising the wavelength difference to the desired rest-frame 
band, thus also minimising colour gradients in the reconstructed PSF. However 
the seeing is wavelength dependent and can also vary on the short timescales 
between the integration in the different SDSS filter bands. This can still 
result in small differences between the actual PSF present for a 
reconstructed galaxy image and the reconstructed PSF image. In order to 
remove this wavelength dependency we choose a well sampled, not saturated 
star in each SDSS filter as PSF. We convolve the SDSS images with a kernel 
(as described above) that generates a round Gaussian PSF with a FWHM for all 
filters of 1.1$\times$ the largest extent of the input PSFs. Thus, PSF 
resolution effects and colour gradients are kept at an absolute minimum. A 
comparison of the reconstructed and target ACS PSF shows a difference of less 
than 0.1\% in each individual pixel.

This procedure is limited to cases where the widths of in- and output PSF are
sufficiently different. If not, noise will introduce artifacts as ringing
patterns or mathematical ghost images near bright sources, which is clearly not
desirable. This occurs primarily when the input PSF resolution becomes
comparable to the output PSF width -- so the convolution function becomes very
narrow -- i.e.~at the low redshift end.

In order to facilitate proper modelling of the redshifted galaxy images we
discourage the use of the publicly available GEMS or COSMOS PSF, but suggest to
prefer the reconstructed PSF instead. Although this might not be crucial for
determining two-dimensional galaxy light profiles, it might make a significant
difference in the case of AGNs and their host galaxies. Likewise, we recommend
using scaled versions of the reconstructed PSF to be put on top of redshifted
galaxy images in order to create artificial AGNs.

Noise in the output images has two main sources, sky background and the galaxy
flux itself. For one-orbit exposures as for COSMOS and GEMS the sky background
noise dominates over the readout noise ($\sim$5.2~$e^-$ for the ACS Wide-Field
Camera) after a few 100 seconds of integration, in the case of broad band
filters. Thus the latter is negligible. Also, at 0.0022~$e^-$~$s^{-1}$ the 
ACS/WFC dark current does not play a significant role.

The background noise in our simulations is not created from random numbers, but
the noise-added galaxy images are added onto blank sky taken from the observed
data itself 
\citep[for ACS data reduction for GEMS and COSMOS see][]{cald06,koek06}. 
In this way reduction signatures like correlated noise and intrinsic small 
variations in the otherwise empty regions of the sky are also present in the 
simulated data. We extract several 60$''$ square fields from the data. While 
we choose comparably empty regions of the sky, the ACS camera is so efficient 
over one orbit for both the COSMOS and GEMS filters that there exists no 
``empty'' sky over more than 10--20$''$ regions. To remove prominent 
remaining objects we replace the corresponding pixel regions with unique other
small patches of blank sky. This process does not involve any filtering, only
replacing. In this way the resulting empty sky fields are very good random
representations of empty sky regions. If for an application the sensitivity 
of a measurement to small residual back-/foreground objects is to be tested, 
the user can add his/her favorite contamination or even use random patches of 
uncleaned sky as background.

Noise has to be added also to the redshifted galaxy images themselves. The S/N
per pixel of the input galaxies is high by definition of the sample. After
redshifting and PSF adaptation their photon noise is negligible compared to the
noise expected from a single orbit HST ACS exposure. The resulting images are
scaled in flux corresponding to e.g.\ 2028 seconds (COSMOS) integration time,
in the respective output filter. The galaxy image -- in units of electrons per
pixel -- then has random Poisson noise added, with a $\sigma^2$ corresponding
to the galaxy flux per pixel. Subsequently, the galaxy is added on top of the
empty background images. In a statistical sense this procedure is not strictly
correct, as the photon noise addition should be applied to the sum of sky and
galaxy, but can not be avoided if real sky images are to be used. However, the
resulting higher noise only appears in regions of the final image where sky
background and galaxy contribute equally in noise, so in the faint isophote
regime of the galaxy.

\tabletypesize{\scriptsize}
\begin{deluxetable*}{c p{3.4cm} p{3.8cm} p{4.2cm} c c}
\tablecaption{Redshifting Test Procedure.\label{tab:tests}}
\tablewidth{0pt}
\tablehead{\colhead{No.} & \colhead{Task} & \colhead{Provides set of} & 
\colhead{Tests} & \colhead{Affected Bands} & \colhead{Fig.}}
\startdata
0 & fit real images using \galfit & \galfit\ parameters & & $u$, $g$, $r$, 
$i$, $z$ & - \\
\hline
1 & from output of 0 create local & artificial SDSS galaxies & 
double-check with 0 & $u$, $g$, $r$, $i$, $z$ & - \\ 
 & smooth model images \& fit & & & & \\ 
\hline
2 & create high-z smooth & smooth model images at high-z & 
\galfit\ response to decreasing S/N & $u$, $g$, $r$ & \ref{fig:ell01} \\ 
 & model images \& fit & & & & \\ 
\hline
3 & redshifting of output from 1 & redshifted, non-$K$-corrected & 
impact of downscaling, PSF effects & $u$, $g$, $r$ & \ref{fig:ell02} \\ 
 & ($K$-corr. disabled) \& fit & smooth model images & & & \\ 
\hline
4 & redshifting of output from 1 & fully redshifted, $K$-corrected & 
impact of $K$-correction & ACS filter & \ref{fig:ell03} \\ 
 & ($K$-corr. enabled) \& fit & smooth model images & & & \\ 
\hline
5 & redshifting of 0 \& fit & redshifted real images & in comparison with 4, & 
ACS filter & \ref{fig:ell06} \\ 
 & & & impact of morphology & & \\ 
\enddata
\end{deluxetable*}
\tabletypesize{\footnotesize}

\section{Resulting galaxy images}\label{sec:results}
Artificially redshifted galaxies are created for a redshift range out to
$z=1.1$. The starting redshift was taken for each galaxy either $z=0.1$ or the
minimum redshift at which both input PSF width and pixel size begin to match 
the output PSF and pixel size. We give images for steps of $\Delta z=0.05$ 
out to $z=0.5$ and $\Delta z=0.1$ beyond that, thus a maximum of 15 output 
redshifts per input galaxy. 

Simply shifting local galaxies out to high redshift makes them look rather
faint in comparison to real average galaxies at such distances. In order to
reflect the brightness increase of high redshift sources we put in a crude
mechanism to introduce evolution. This optional feature allows to make galaxies
brighter as a linear function of redshift:
\[
M_{\mathrm{evo}}=x\times z + M
\]
Setting $x=-1$ would make a galaxy 1 mag brighter at redshift $z=1$ as it would
normally be. Yet, this option is not meant to be a substitute for real
morphological or photometrical evolution and does not replace stellar evolution
codes. The reason for putting in such a simple functional form is the
application of galaxy classification by eye. If one is to re-identify galaxies
shifted to high redshift, the task is made increasingly unfair compared to real
galaxies, which on average do become brighter at higher redshift, if one does
not apply artificial brightening.

A few resulting example images for COSMOS are shown in 
Figure~\ref{fig:examples}. All images, for both GEMS and COSMOS are made 
available electronically.

\section{Tests of \ferengi\ and application of GALFIT}\label{sec:tests}
In order to demonstrate the accuracy of \ferengi\ and characterise its limits, 
we perform a number of tests. We use the programme \galfit\ \citep{peng02} to 
determine structural parameters, which is often used for parameter estimation 
and morphological classification of galaxy samples, particularly in survey 
applications. The programme is (potentially) susceptible to S/N changes and 
morphological $K$-correction, like any other fitting code. Yet, 
\citet{haus07} have shown that \galfit\ is very robust and does not exhibit 
systematic biases when fitting radial surface brightness models to artificial 
2d \sersic-type\footnote{The \sersic\ profile is a generalized exponential 
profile, with one parameter $n$. For $n=1$ the profile is a simple 
exponential curve, typical of disc galaxies, for $n=4$ the profile becomes a 
de~Vaucouleurs exponential $r^{1/4}$ profile.} light profiles. In the 
following section we make use of this robustness and apply \galfit\ to 
various sets of simulated 2d \sersic\ profiles to determine the dependence of 
structural parameters on the application of \ferengi.

In addition to \galfit\ we use a code described in \citet{haus07} to create 2d
\sersic\ profiles mimicking our observed galaxies. This procedure features a 
fine subsampling of the inner pixels allowing highly accurate flux calibration.
Convolution is done with the original PSF as reconstructed from the SDSS,
after smoothing the individual five bands to a round Gausssian configuration.
Poisson noise and a real SDSS sky background image are added. Such simulated
images are created for all five SDSS bands $u$, $g$, $r$, $i$, $z$.

\subsection{Procedure for individual galaxies}

\begin{figure*}[t]
\plotone{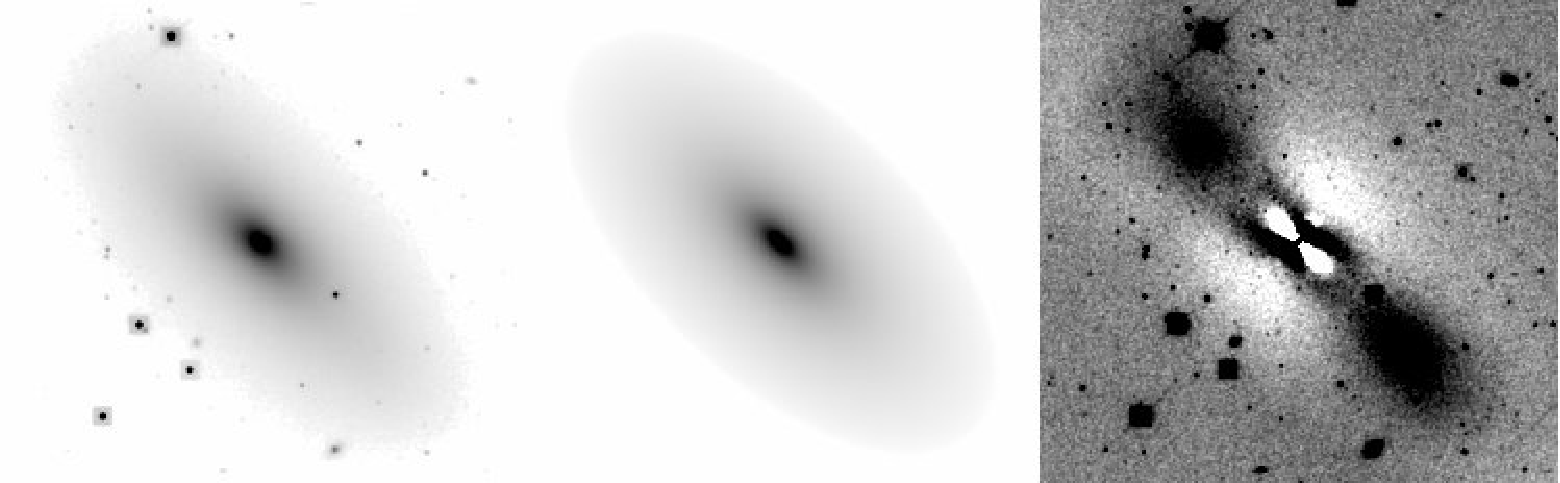}
\caption{\label{fig:example_ell} 
Example galaxy for testing \ferengi. Original SDSS $r$-band image (left); 
\galfit\ model image (middle); residual image (SDSS minus \galfit\ model; 
right). The differences in the residual image are $\sim 15\%$ in the inner 
and 2-3\% in the outer regions (with respect to the flux in the original 
image).
}
\end{figure*}

We conduct the simulations in six different steps to check for possible 
systematics introduced at each step. The summary of each step is shown in 
Table~\ref{tab:tests}. We show by way of illustration an example of an 
elliptical galaxy (Fig.~\ref{fig:example_ell}). It has an almost de 
Vaucouleurs-type light profile ($n$ increases from 3.2 to 3.6 going from $u$ 
to $r$ band). The size of the galaxy is constant across the three bands; the 
magnitude covers a range of over 2.5 magnitudes -- $u$ being faintest and $r$ 
brightest, as expected for a red elliptical.

$\bullet$ We start by using \galfit\ to fit a \sersic\ profile to the set of
SDSS images belonging to the objects. This provides the basis for simulating 
the smooth model images.

$\bullet$ With the \galfit\ parameters we create artificial model images 
using the procedure by \citet{haus07} described above, resembling smooth 
versions of the SDSS galaxy images. A fit to these smooth images in all bands 
confirms the stability of the process by retrieving almost exactly the input 
values.

$\bullet$ Next we create artificial model images of the same galaxies, as 
they would appear at high redshift (our set of output redshifts given in 
Section~\ref{sec:results}) using the parameters from before, in the $u$, $g$, 
and $r$ bands. We do not apply \ferengi\ here, but only convert the linear 
scales according to redshift. This will quantify the response of \galfit\ to 
the decreasing S/N as a result of cosmological surface brightness dimming.

\begin{figure}[b]
\plotone{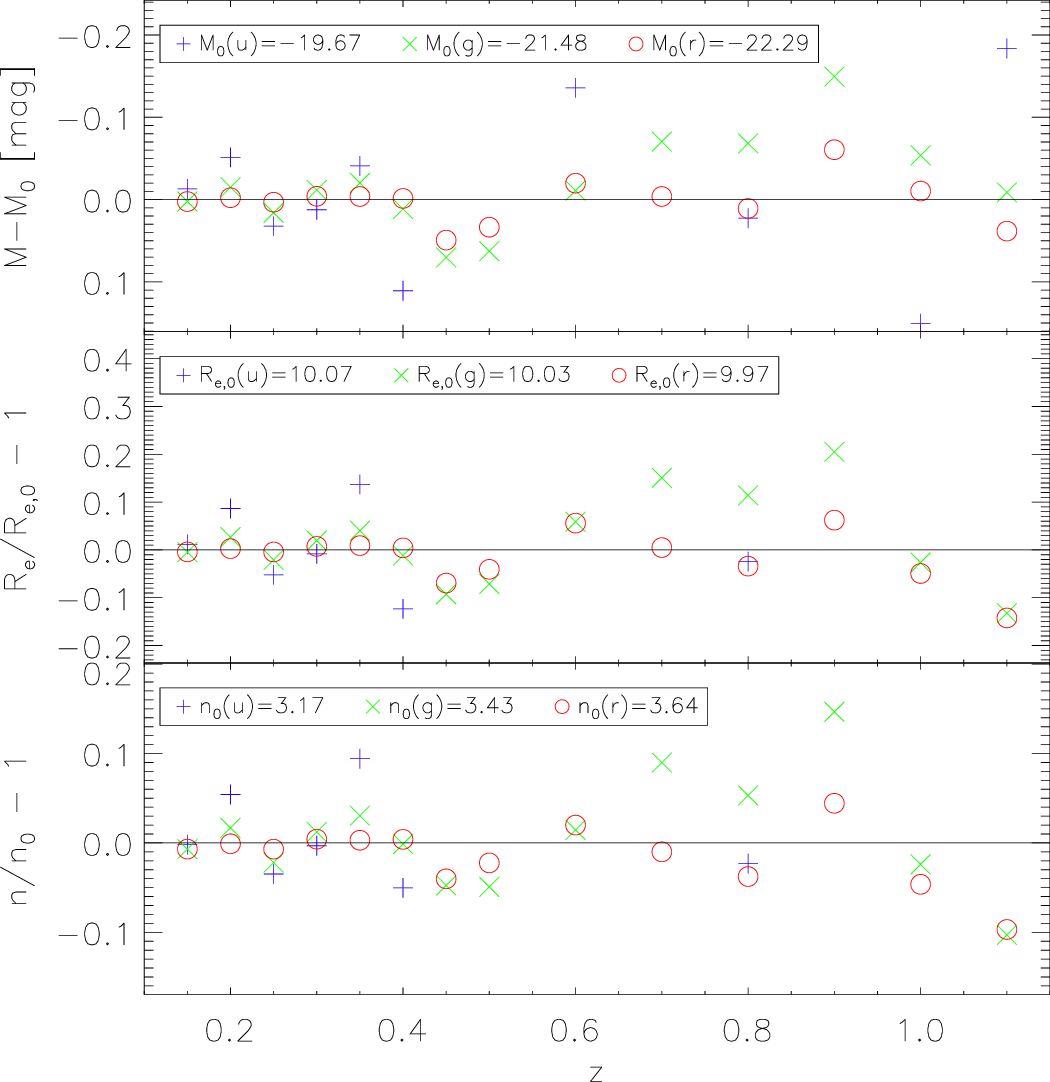}
\caption{\label{fig:ell01} 
The \galfit\ error from fitting simulated images for an example galaxy. Top 
panel: absolute magnitude difference; middle panel: ratio of measured and 
input half-light radius; bottom panel: ratio of measured and input \sersic\ 
index. Pluses, crosses and circles show $u$-, $g$- and $r$-band, 
respectively. Boxes in each panel indicate the absolute quantities of the 
simulation input: absolute magnitude, physical half-light radius in kpc and 
\sersic\ index.
}
\end{figure}

Figure~\ref{fig:ell01} demonstrates how \galfit\ performs as a function of
redshift for the example galaxy (No.~2 from Table~\ref{tab:tests}). The higher
the redshift, the lower is the S/N of the object and its surface brightness.
Quite expectedly, the quality of the recovered fit parameters degrades with
redshift. As the $u$-band has the lowest S/N of all SDSS filters, \galfit\ 
fares worst there; $r$-band behaves best, having the highest S/N. The chosen 
galaxy being rather red, this effect is even amplified. As the galaxy is 
rather bright, overall, the deviations in the parameters from their input 
values are rather small. In order to allow a comparison with GEMS, we compute 
the average apparent surface brightness of our sample galaxies with $3<n<5$ 
in $u$, $g$ and $r$. From \citet{haus07} we obtain the corresponding \galfit\ 
errors in GEMS. The results are listed in Table~\ref{tab:gemserrors}.

\begin{table}[b]
\begin{center}
\caption{Statistical Errors from GEMS.\label{tab:gemserrors}}
\begin{tabular}{ccccc}
\tableline
\tableline
z & Band & $M$ & $R_e$ & $n$ \\
  &      & [mag] & [\%] & [\%] \\
\tableline
    & $u$ & 0.05 $\pm$ 0.42 & -0.06 $\pm$ 0.41 & -0.03 $\pm$ 0.27 \\
0.2 & $g$ & 0.01 $\pm$ 0.15 &  0.00 $\pm$ 0.22 & -0.01 $\pm$ 0.21 \\
\vspace{0.5mm}
    & $r$ & 0.01 $\pm$ 0.13 &  0.00 $\pm$ 0.20 & -0.01 $\pm$ 0.20 \\
    & $u$ & 0.10 $\pm$ 0.50 & -0.08 $\pm$ 0.48 & -0.05 $\pm$ 0.28 \\
0.4 & $g$ & 0.02 $\pm$ 0.20 & -0.01 $\pm$ 0.27 & -0.01 $\pm$ 0.22 \\
\vspace{0.5mm}
    & $r$ & 0.01 $\pm$ 0.17 &  0.00 $\pm$ 0.24 & -0.01 $\pm$ 0.21 \\
    & $u$ & 0.18 $\pm$ 0.61 & -0.15 $\pm$ 0.50 & -0.06 $\pm$ 0.31 \\
0.8 & $g$ & 0.03 $\pm$ 0.32 & -0.04 $\pm$ 0.35 & -0.02 $\pm$ 0.26 \\
    & $r$ & 0.02 $\pm$ 0.27 & -0.03 $\pm$ 0.32 & -0.02 $\pm$ 0.25 \\
\tableline
\end{tabular}
\end{center}
\tablenotetext{}{The morphological parameters of the average galaxy
with $3<n<5$ in our sample are: $M$=[-19.53, -21.09, -21.95], $R_e$=[12.0, 7.7,
9.9] in [$u$, $g$, $r$]. Indicated errors are $1\sigma$ standard deviations 
from the mean.}
\end{table}

\begin{figure}[t]
\plotone{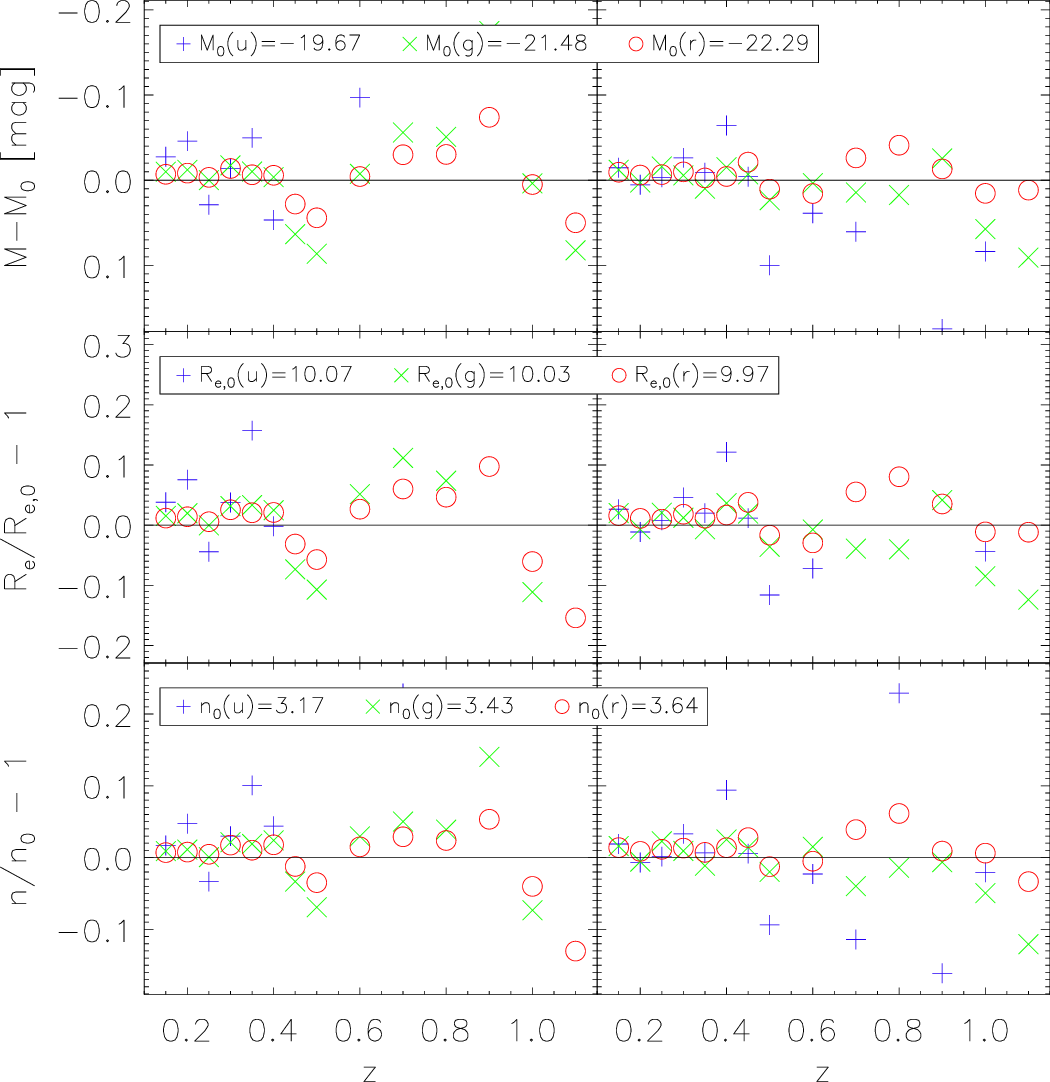}
\caption{\label{fig:ell02} 
Performance of \ferengi\ (for an example galaxy) -- $K$-correction 
{\it disabled}. Same axes as in Figure~\ref{fig:ell01}. Left panel: difference 
of simulated parameters and output after redshifting without applying 
$K$-corrections. Right panel: same as left panel minus the offsets from 
Figure~\ref{fig:ell01}, thus indicating the additional error implied by 
re-gridding, PSF-resampling and changed noise properties.
}
\end{figure}

\begin{figure*}[t]
\includegraphics[clip,height=5cm]{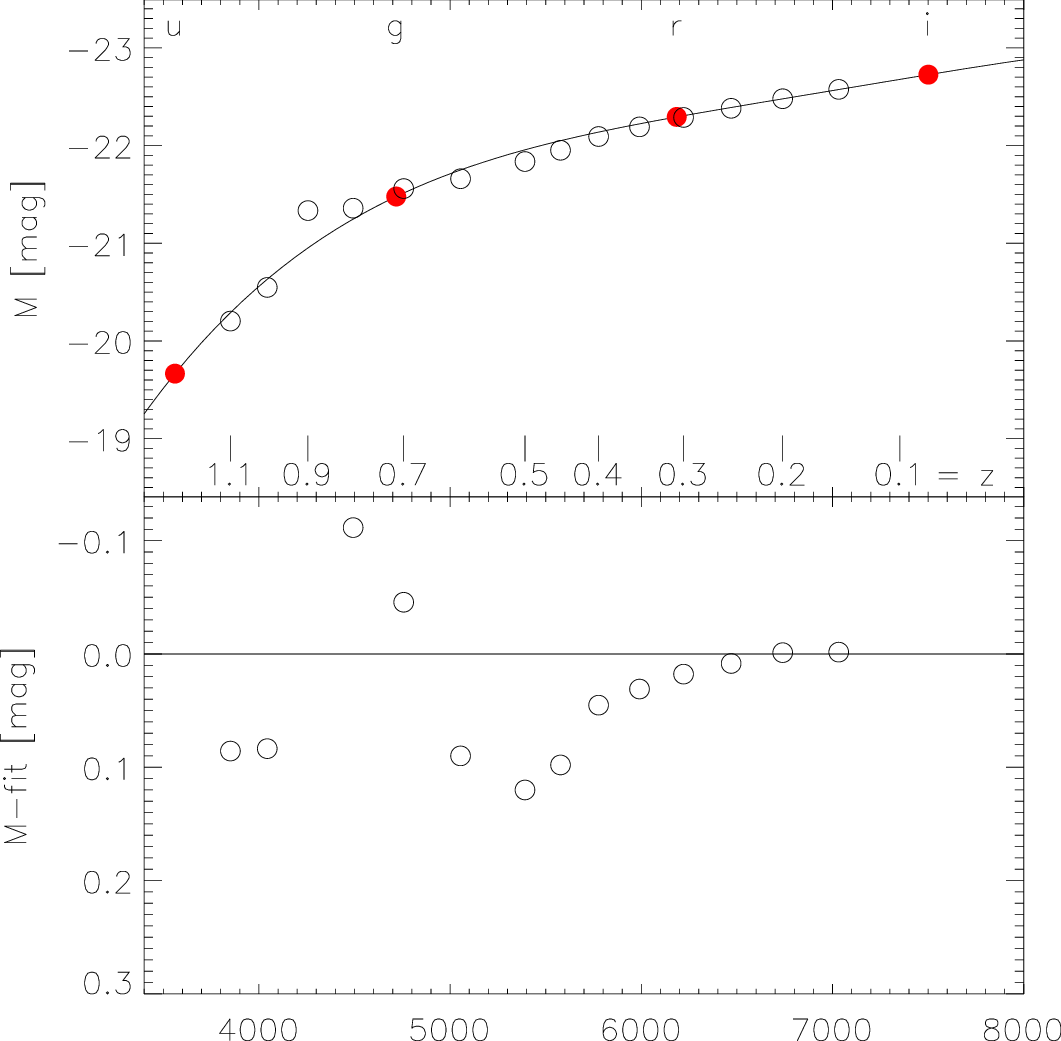}%
\hfill\includegraphics[clip,height=5cm]{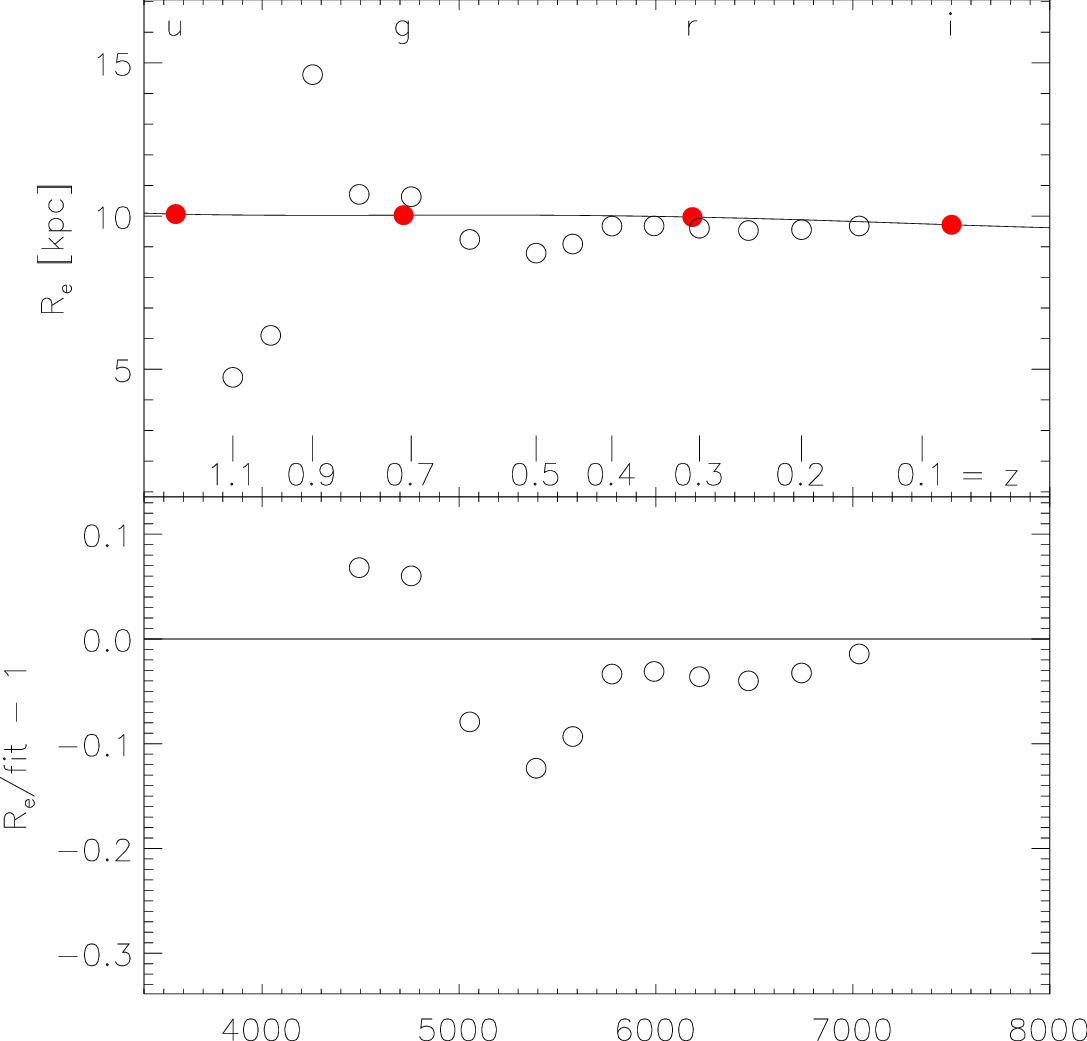}%
\hfill\includegraphics[clip,height=5cm]{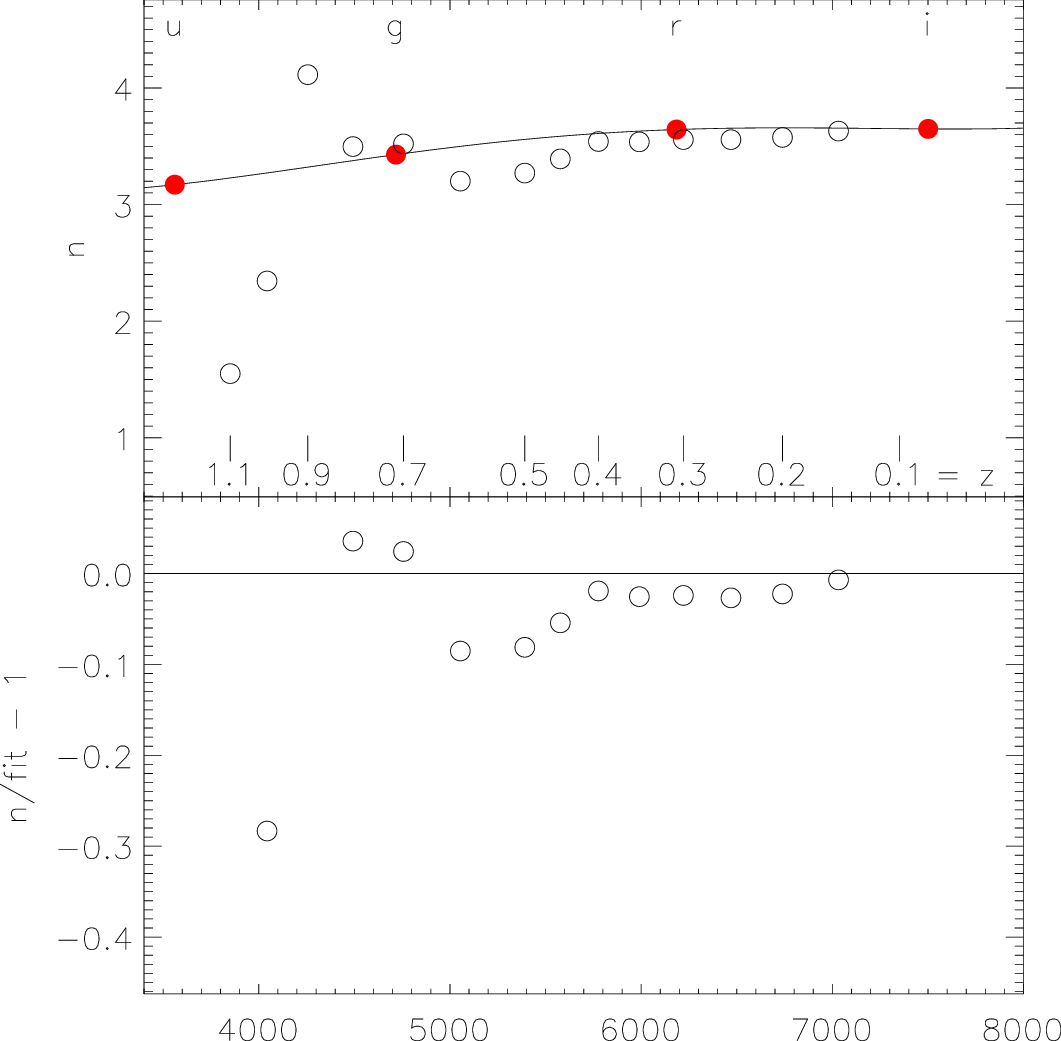}
\caption{\label{fig:ell03} 
Performance of \ferengi\ (for an example galaxy) -- $K$-correction 
{\it enabled}. Top panels: Turning on $K$-corrections we compare the absolute 
magnitudes, half-light radii and \sersic\ indices of the simulated SDSS 
images (filled circles), which were used as input for the redshifting, 
directly with the redshifted output (open circles). The solid line marks a 
polynomial fit to the input data. Bottom panel: subtracting the polynomial 
fit from the output data.
}
\end{figure*}

$\bullet$ Now we use \ferengi\ on the simulated smooth $u$, $g$, and $r$ band 
model images to create versions of these images appropriate for higher 
redshifts. The \sersic\ fits to this set of images we use to determine the 
influence of conversion to a new pixel grid and related PSF on different 
galaxy parameters, as a function of redshift. Note that the $K$-correction 
code is {\it disabled} for this purpose. The image set contains the flux of the 
SDSS filters transformed to a new pixel size, PSF, background noise 
properties and the cosmological effect of surface brightness dimming. 

For the example galaxy the left hand side of Figure~\ref{fig:ell02} shows the 
same as Figure~\ref{fig:ell01}; the right hand side has the values from 
Figure~\ref{fig:ell01} subtracted in order to indicate the extra influence of 
the re-gridding process. We find no additional systematics and the scatter 
hardly increases.

$\bullet$ After that we {\it enable} $K$-correction, thus fitting an SED
to all five SDSS bands and extracting the flux at the position of the
observed ACS filter, in addition to the down-scaling and surface
brightness dimming. This provides a set of images representing the
simulated SDSS galaxies as they would look at high redshift including all
cosmological effects (No.~4 from Table~\ref{tab:tests}).

The result for the example galaxy is shown in Figure~\ref{fig:ell03}. In the 
top panels we plot the original input values. Also shown are the fit values 
from using the $K$-correction code. In order to provide a rough estimate of 
the deviation from the expected values we fit the five values from the 
simulated SDSS images with a polynomial and subtract this fit. Going into all 
the details of modelling galaxy photometry and structure in multiple 
wavebands (or even continuously) is well beyond the scope of this paper. 
Therefore, we regard the lower panel not as a proper estimate of the error 
budget, but rather an indication of systematic trends. Points missing from 
the lower panel were left out in order not to overstretch the plotting axis 
and to focus on the main region of interest.

\begin{figure*}[b]
\hfill\includegraphics[clip,height=8cm]{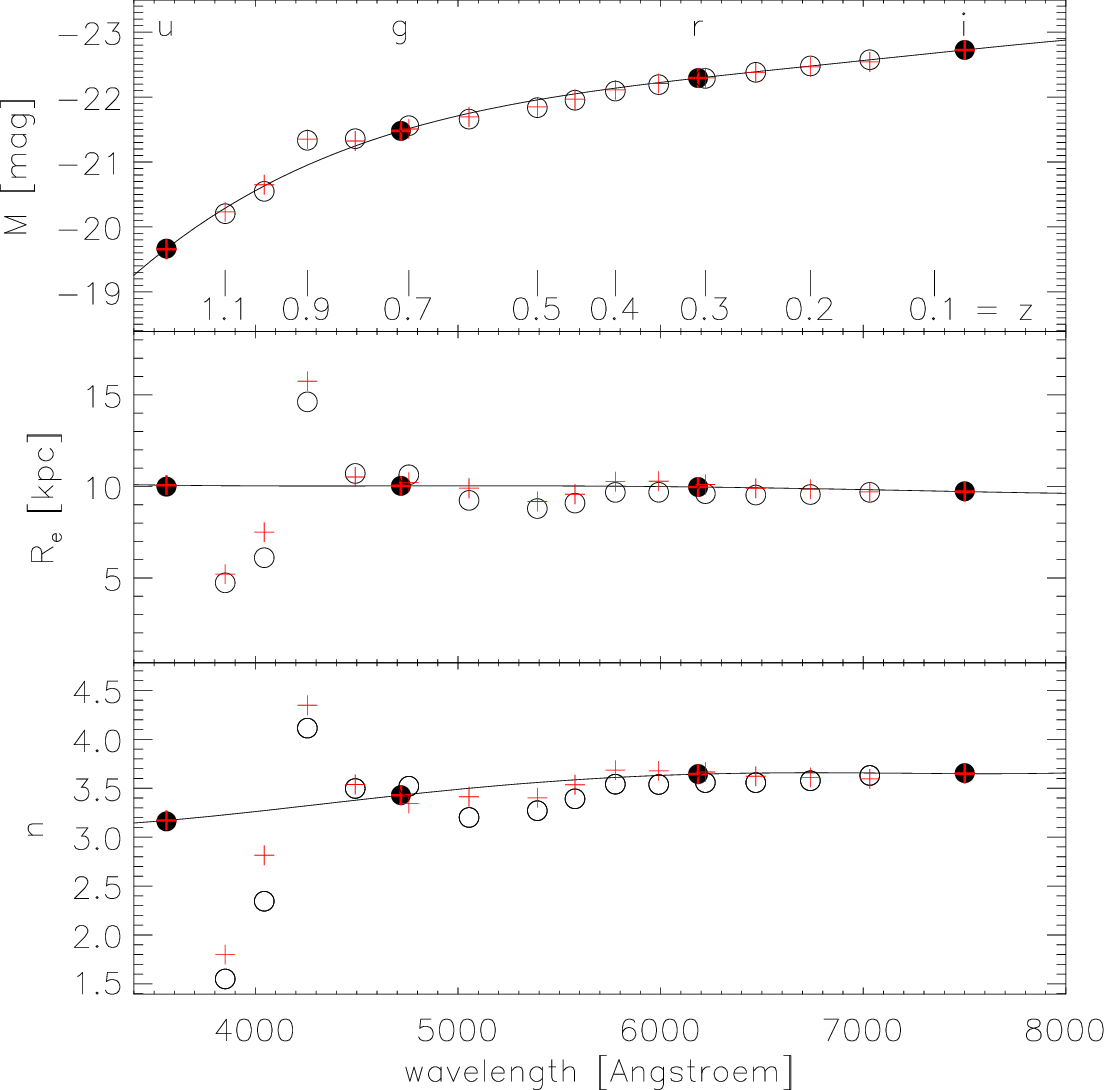}%
\hfill\includegraphics[clip,height=8cm]{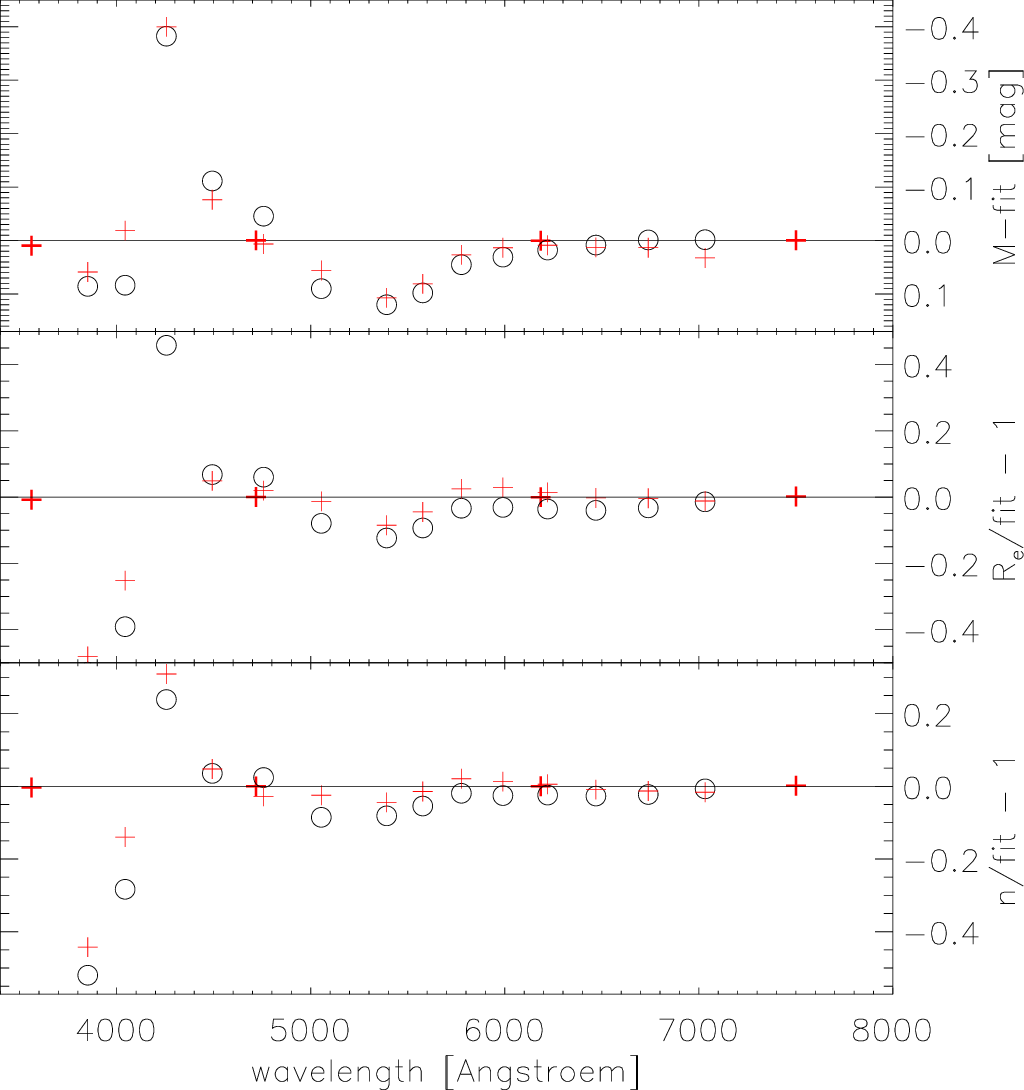}\hspace*{\fill}
\caption{\label{fig:ell06}
Comparison of reality and simulation (for an example galaxy). Plots for 
magnitude, half-light radius and \sersic\ index from top to bottom. Simulated 
data shown as circles; real data as pluses. Thick / filled symbols indicate 
local input data. Left: absolute quantities. The solid line shows a 
polynomial fit to the simulated data. Right: polynomial fit subtracted from 
the data. Hardly any difference is visible, indicating that the input galaxy 
fits well the featureless \sersic\ profile.
}
\end{figure*}

$\bullet$ Finally, we run \ferengi\ on the real SDSS images including 
$K$-corrections (No.~5 from Table~\ref{tab:tests}). The \galfit\ results for 
this series reveals in comparison to the redshifted simulations (No.~4 from 
Table~\ref{tab:tests}) the impact of morphology on the whole process (see 
Figure~\ref{fig:ell06}).

As the performed calculations are virtually identical any differences in the 
output must result from irregularities in the real data, such as tidal 
features, spiral arms, dust lanes, etc. The example galaxy was chosen to be 
rather featureless. Therefore, a perfect match is obtained. Yet, in the 
majority of cases we find systematic differences: Either redshift dependent 
deviations (originating from irregularities) or global shifts towards higher 
or lower values. At the highest redshift ($z\sim 1$) the impact of morphology 
might decrease again as virtually all features are smeared out.

Yet, the reason for global offsets is of technical nature: the drastic change 
in resolution and depth when transforming a low redshift image ($cz$ of a few 
thousand km/h) to high redshift ($z\gtrsim 0.1$) results in the rebinned PSF 
not being Nyquist sampled any more. This causes shifts when performing 
Fourier transformations. The effect is more pronounced at low 
($z\gtrsim 0.1$) than at high ($z\sim 1$) redshift, because eventually the 
PSF becomes pointlike.

\subsection{Average results}

\begin{figure}[htp]
\plotone{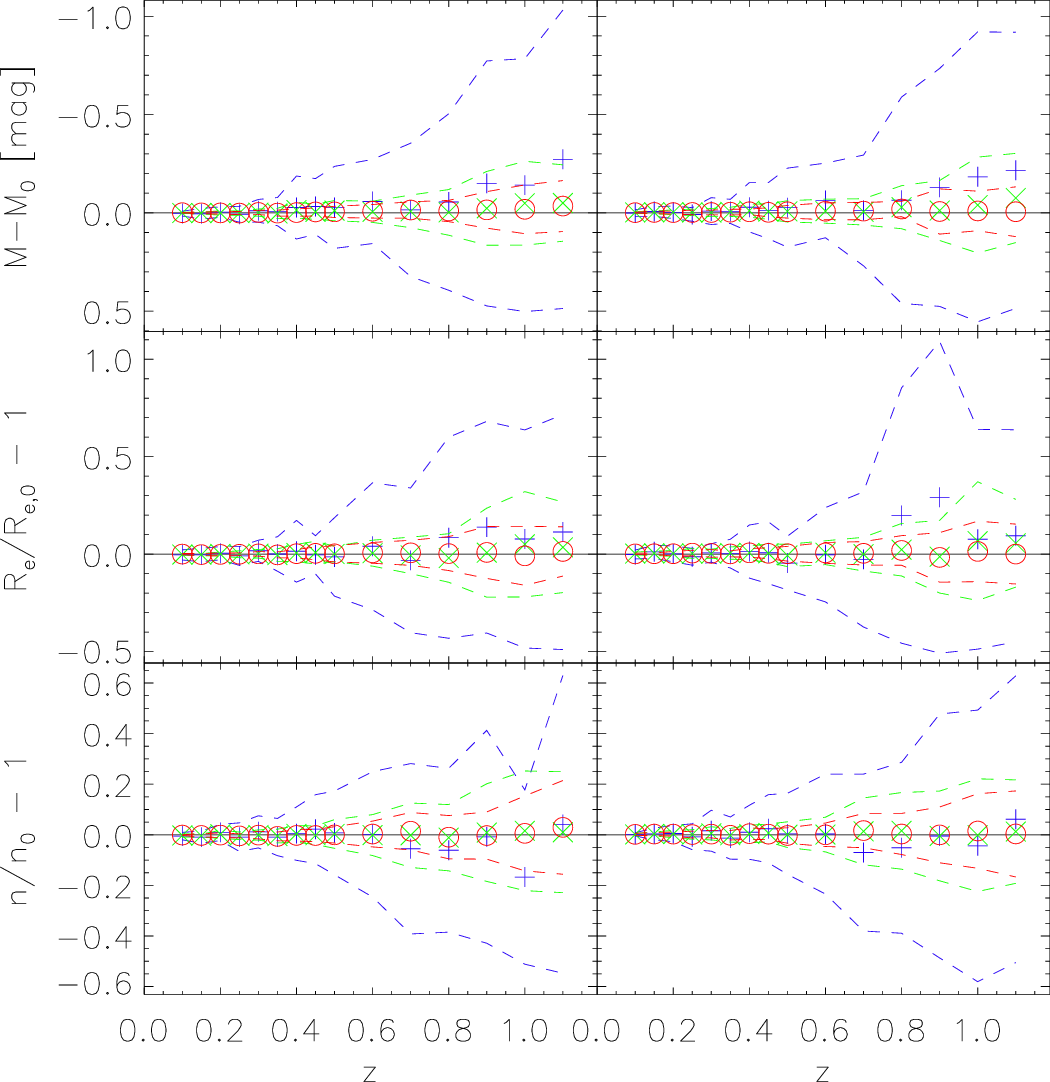}
\caption{\label{fig:mean01}
Average deviations from the input values (mean values for the whole sample). 
Left: simulating synthetic galaxies at various redshifts. Right: applying 
\ferengi\ to simulated SDSS galaxies without $K$-correction. From top to 
bottom: absolute magnitude difference, ratio of measured and input half-light 
radius, ratio of measured and input \sersic\ index. Pluses, crosses and 
circles symbolise $u$-, $g$- and $r$-band, respectively. The error bars are 
robust $1\sigma$ standard deviations from the mean.
}
\end{figure}

We applied \ferengi\ to all \Ngal\ sample galaxies. This allows us to 
quantify systematic biases of the redshifting code in some detail. In 
Figure~\ref{fig:mean01} we show the average deviation from the input values 
for simulations of high redshift galaxies and downscaled versions of simulated 
galaxies (No.~2 \& 3 from Table~\ref{tab:tests}). The errors and systematic 
offsets are not significantly different. For the given observational setup 
(redshifting of SDSS images to COSMOS and GEMS) roughly at $z\sim0.7$ larger 
deviations in particular in $u$-band occur. Note, that these depatures from 
the expected mean are seen in the simulated images as well. Therefore, they 
must originate from the specific \galfit\ setup, but not \ferengi\ itself.

\begin{figure*}[htp]
\plotone{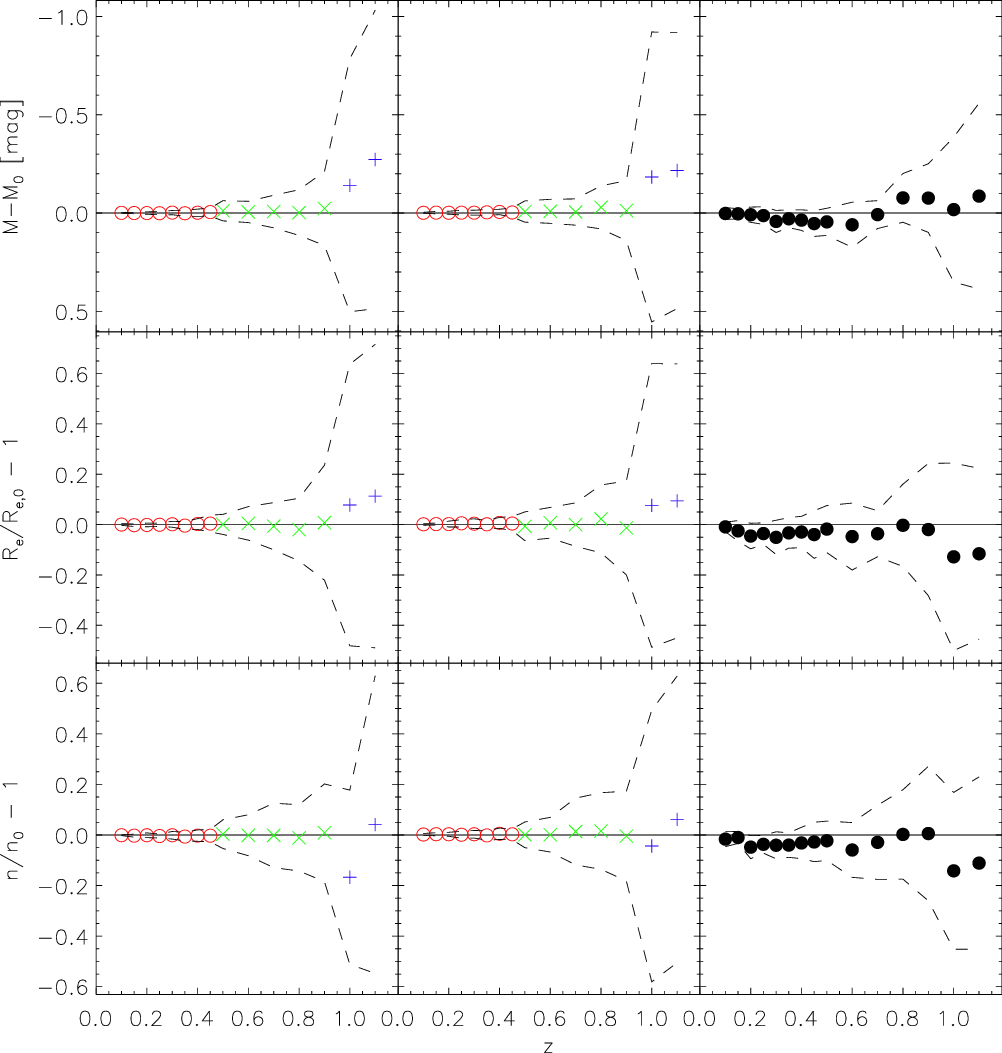}
\caption{\label{fig:mean02}
Average deviations from the input values including $K$-corrections. Panel lines
show absolute magnitude difference, ratio of measured and input half-light
radius and \sersic\ index; panel columns show simulation of high redshift
galaxies, redshifting of simulated galaxies without $K$-corrections and
redshifting of simulated galaxies with $K$-correction enabled
(No.~2, 3 \& 4 of Table~\ref{tab:tests}, respectively). The error bars are
robust $1\sigma$ standard deviations from the mean. The leftmost two panel
columns show data from Figure~\ref{fig:mean01}. In contrast to
Figure~\ref{fig:mean01}, we choose to plot at each redshift the SDSS band
closest to the rest-frame of the ACS filter ($z\geq1$: $u$-band -- pluses;
$0.5\leq z\leq0.9$: $g$-band -- crosses; $z\leq0.45$: $r$-band -- circles).
}
\end{figure*}

In order to compare the results of the high redshift simulations and the
downscaling-only simulations (No.~2 \& 3 from Table~\ref{tab:tests}) with the
$K$-corrections-enabled setup (No.~4), we plot as a function of redshift the
measurements in the SDSS filters closest to the rest-frame ACS filter
(Figure~\ref{fig:mean02}). Interestingly, the error bars in the right panel of
Figure~\ref{fig:mean02} (No.~4) at the highest two redshift bins are {\it
smaller} than without $K$-corrections (No.~3) or even in the pure simulations
(No.~2). The reason for this is, that the SED fitting uses all five SDSS bands
and therefore introduces some information extrapolated from $g$, $r$, $i$ and
$z$ (the closer bands being weighted more strongly than the redder bands) to
improve the low S/N $u$-band data. However, this implies also that
template mismatches might change the morphological appearance of the
object to some extent. As the SED-fitting code is deeply embedded in
\kcorrect\ we do not attempt to characterise the impact any further. Moreover,
we find, that the $K$-correction code on average introduces stronger
fluctuations and increases the error bars slightly at lower redshifts.
Systematic deviations to lower brightnesses, radii or \sersic\ indices are not
statistically significant.

\section{Discussion and conclusions}\label{sec:conclusions}
We take from Figures~\ref{fig:mean01} and \ref{fig:mean02} that \ferengi\ has 
no unwanted systematic effects on measured morphological and photometrical 
parameters of simulated galaxies. The deviations from theoretical values are 
well within the uncertainties expected from, in this case, \galfit, also in 
absolute terms. Both linear scales as well as magnitudes are well reproduced. 
This includes the bandpass shifts induced by redshifting and different 
filters, as is demonstrated in the right panels in Figures~\ref{fig:mean02}: 
The \kcorrect\ module correctly interpolates between input filter images. We 
cannot exclude issues when {\em extrapolating} \kcorrect\ templates, but we 
explicitly restrict the output range of redshifts to only use template 
{\em interpolation}, between bands.

We make \ferengi\ publically available from our webpage, as are redshifted 
sets of images for redshifts $0.1\le z\le1.1$ and different HST ACS filters. 
Primarily these are created for the GEMS (F606W and F850LP at 
0\farcs03/pixel) and COSMOS (F814W at 0\farcs05/pixel) projects, but the code 
can be modified and used for other filters.
The user can extend the input sample of images (also provided on the
webpage), to other objects, filter curves, and output redshifts, and in
principle also to other data sources. We leave it up to the user to update 
\ferengi\ with respect to the $K$-correction module when including other input
filter bands.

As stated, templates are used for interpolation between input filters. They
can theoretically also be extrapolated beyond the input wavelength interval,
if one has faith in the templates. However, we strongly advise against this
for a different reason: Due to the limited S/N available for the
pixel-by-pixel $K$-correction that is computed, a mild extrapolation might be
acceptable, but the noise of the output pixels and systematic deviations will 
obviously increase with distance to the last supported wavelength. As various 
templates might be degenerate when interpolating, there might be striking 
differences on extrapolation, resulting in large errors in the output flux.

As a second caveat we note that stars in the input images are not treated
separately in \ferengi\ or in the provided datasets. {\it Any} flux in the
input images, be it stars or back-/foreground galaxies, will be treated as if
it were at the distance of the target and thus be redshifted. While distant
background galaxies usually disappear, stars will be represented with the
output PSF, resembling a faint dense star field (the density of stars is
enhanced quadratically with decreasing linear scales as the simulation
redshift increases). Of even greater concern are foreground (or closer
background) galaxies. In the output, their physical properties are calculated
falsely as their assumed distance is not correct. Whether the resulting images
will in the end be reliable representations of reality, strongly depends on
the quality of input images, the type of galaxies, and their correspondence
with the templates used.

\acknowledgements
The authors gratefully acknowledge numerous helpful discussions with Chien Y. 
Peng and Hans-Walter Rix.
This publication makes use of the Sloan Digital Sky Survey, 
\url{http://www.sdss.org}. We selected the SDSS data using the HyperLeda 
database (\url{http://leda.univ-lyon1.fr/intro.html}).
This work is also based on observations taken with the NASA/ESA 
{\it Hubble Space Telescope}, which is operated by AURA. Support for the GEMS 
project (\url{http://www.mpia.de/GEMS/gems.htm}) was provided by NASA through 
grant number HST-GO-9500. The HST COSMOS Treasury program was supported 
through NASA grant HST-GO-09822 (\url{http://cosmos.astro.caltech.edu}).
M.B.\ was supported in part by the Austrian Science Foundation FWF under 
grant P18416.
K.J.\ was supported by the German DLR under project number 50~OR~0404 and
by the German DFG with grant SCHI 536/3-1 within the priority program
1177.


\begin{thebibliography}{}

\bibitem[{{Abazajian} {et~al.}(2005){Abazajian}, {Adelman-McCarthy}, {Ag{\"
  u}eros}, {Allam}, {Anderson}, {Anderson}, {Annis}, {Bahcall}, {Baldry},
  {et~al.}}]{abaz05}
{Abazajian}, K., {Adelman-McCarthy}, J.~K., {Ag{\" u}eros}, M.~A., {Allam},
  S.~S., {Anderson}, K.~S.~J., {Anderson}, S.~F., {Annis}, J., {Bahcall},
  N.~A., {Baldry}, I.~K., {et~al.} 2005, AJ, 129, 1755

\bibitem[{{Abraham} {et~al.}(1996{\natexlab{a}}){Abraham}, {Tanvir},
  {Santiago}, {Ellis}, {Glazebrook}, \& {van den Bergh}}]{abraham1996a}
{Abraham}, R.~G., {Tanvir}, N.~R., {Santiago}, B.~X., {Ellis}, R.~S.,
  {Glazebrook}, K., \& {van den Bergh}, S. 1996{\natexlab{a}}, \mnras, 279, L47

\bibitem[{{Abraham} {et~al.}(1996{\natexlab{b}}){Abraham}, {van den Bergh},
  {Glazebrook}, {Ellis}, {Santiago}, {Surma}, \& {Griffiths}}]{abraham1996b}
{Abraham}, R.~G., {van den Bergh}, S., {Glazebrook}, K., {Ellis}, R.~S.,
  {Santiago}, B.~X., {Surma}, P., \& {Griffiths}, R.~E. 1996{\natexlab{b}},
  \apjs, 107, 1

\bibitem[{{Barden} {et~al.}(2005){Barden}, {Rix}, {Somerville}, {Bell},
  {H{\"a}u{\ss}ler}, {Peng}, {Borch}, {Beckwith}, {Caldwell}, {Heymans},
  {Jahnke}, {Jogee}, {McIntosh}, {Meisenheimer}, {S{\'a}nchez}, {Wisotzki}, \&
  {Wolf}}]{barden2005}
{Barden}, M., {Rix}, H.-W., {Somerville}, R.~S., {Bell}, E.~F.,
  {H{\"a}u{\ss}ler}, B., {Peng}, C.~Y., {Borch}, A., {Beckwith}, S.~V.~W.,
  {Caldwell}, J.~A.~R., {Heymans}, C., {Jahnke}, K., {Jogee}, S., {McIntosh},
  D.~H., {Meisenheimer}, K., {S{\'a}nchez}, S.~F., {Wisotzki}, L., \& {Wolf},
  C. 2005, \apj, 635, 959

\bibitem[{{Blanton} {et~al.}(2003){Blanton}, {Brinkmann}, {Csabai}, {Doi},
  {Eisenstein}, {Fukugita}, {Gunn}, {Hogg}, \& {Schlegel}}]{blan03}
{Blanton}, M.~R., {Brinkmann}, J., {Csabai}, I., {Doi}, M., {Eisenstein}, D.,
  {Fukugita}, M., {Gunn}, J.~E., {Hogg}, D.~W., \& {Schlegel}, D.~J. 2003, AJ,
  125, 2348

\bibitem[{{Bouwens} {et~al.}(1998){Bouwens}, {Broadhurst}, \&
  {Silk}}]{bouwens1998}
{Bouwens}, R., {Broadhurst}, T., \& {Silk}, J. 1998, \apj, 506, 557

\bibitem[{{Bruzual} \& {Charlot}(2003)}]{bc03}
{Bruzual}, G. \& {Charlot}, S. 2003, \mnras, 344, 1000

\bibitem[{{Burgarella} {et~al.}(2001){Burgarella}, {Buat}, {Donas}, {Milliard},
  \& {Chapelon}}]{burgarella2001}
{Burgarella}, D., {Buat}, V., {Donas}, J., {Milliard}, B., \& {Chapelon}, S.
  2001, \aap, 369, 421

\bibitem[{Caldwell {et~al.}(2006)}]{cald06}
Caldwell, J. {et~al.} 2006, PASP, in prep.

\bibitem[{{de Vaucouleurs} {et~al.}(1991){de Vaucouleurs}, {de Vaucouleurs},
  {Corwin}, {Buta}, {Paturel}, \& {Fouque}}]{deva91}
{de Vaucouleurs}, G., {de Vaucouleurs}, A., {Corwin}, H.~G., {Buta}, R.~J.,
  {Paturel}, G., \& {Fouque}, P. 1991, {Third Reference Catalogue of Bright
  Galaxies} (Volume 1-3, XII, 2069 pp.~7 figs..~ Springer-Verlag Berlin
  Heidelberg New York)

\bibitem[{{Giavalisco} {et~al.}(1996){Giavalisco}, {Livio}, {Bohlin},
  {Macchetto}, \& {Stecher}}]{giavalisco1996}
{Giavalisco}, M., {Livio}, M., {Bohlin}, R.~C., {Macchetto}, F.~D., \&
  {Stecher}, T.~P. 1996, \aj, 112, 369

\bibitem[{Gray \& STAGES(2007)}]{grayinprep}
Gray, M. \& STAGES. 2007, MNRAS, in prep.

\bibitem[{H\"au{\ss}ler {et~al.}(2007)}]{haus07}
H\"au{\ss}ler, B. {et~al.} 2007, submitted to ApJ

\bibitem[{Koekemoer {et~al.}(2006)}]{koek06}
Koekemoer, A. {et~al.} 2006, submitted to ApJS

\bibitem[{{Kuchinski} {et~al.}(2001){Kuchinski}, {Madore}, {Freedman}, \&
  {Trewhella}}]{kuchinski2001}
{Kuchinski}, L.~E., {Madore}, B.~F., {Freedman}, W.~L., \& {Trewhella}, M.
  2001, \aj, 122, 729

\bibitem[{{Lilly} {et~al.}(1998){Lilly}, {Schade}, {Ellis}, {Le Fevre},
  {Brinchmann}, {Tresse}, {Abraham}, {Hammer}, {Crampton}, {Colless},
  {Glazebrook}, {Mallen-Ornelas}, \& {Broadhurst}}]{lilly1998}
{Lilly}, S., {Schade}, D., {Ellis}, R., {Le Fevre}, O., {Brinchmann}, J.,
  {Tresse}, L., {Abraham}, R., {Hammer}, F., {Crampton}, D., {Colless}, M.,
  {Glazebrook}, K., {Mallen-Ornelas}, G., \& {Broadhurst}, T. 1998, \apj, 500,
  75

\bibitem[{{Lisker} {et~al.}(2006){Lisker}, {Debattista}, {Ferreras}, \&
  {Erwin}}]{lisker2006}
{Lisker}, T., {Debattista}, V.~P., {Ferreras}, I., \& {Erwin}, P. 2006, \mnras,
  370, 477

\bibitem[{{Paturel} {et~al.}(1989){Paturel}, {Fouque}, {Bottinelli}, \&
  {Gouguenheim}}]{Patu89}
{Paturel}, G., {Fouque}, P., {Bottinelli}, L., \& {Gouguenheim}, L. 1989,
  \aaps, 80, 299

\bibitem[{Peng {et~al.}(2002)Peng, Ho, Impey, \& Rix}]{peng02}
Peng, C.~Y., Ho, L.~C., Impey, C.~D., \& Rix, H.-W. 2002, AJ, 124, 266

\bibitem[{{Prugniel} \& {Heraudeau}(1998)}]{Pru98}
{Prugniel}, P. \& {Heraudeau}, P. 1998, \aaps, 128, 299

\bibitem[{Rix {et~al.}(2004)Rix, Barden, Beckwith, Bell, Borch, Caldwell,
  H\"au{\ss}ler, Jahnke, Jogee, McIntosh, Meisenheimer, Peng, S\'anchez,
  Somerville, Wisotzki, \& Wolf}]{rix04}
Rix, H.-W., Barden, M., Beckwith, S. V.~W., Bell, E.~F., Borch, A., Caldwell,
  J. A.~R., H\"au{\ss}ler, B., Jahnke, K., Jogee, S., McIntosh, D.~H.,
  Meisenheimer, K., Peng, C.~Y., S\'anchez, S.~F., Somerville, R.~S., Wisotzki,
  L., \& Wolf, C. 2004, ApJS, 152, 163

\bibitem[{Scoville {et~al.}(2006)}]{scov06}
Scoville, N. {et~al.} 2006, submitted to ApJS

\bibitem[{{Takamiya}(1999)}]{takamiya1999}
{Takamiya}, M. 1999, \apjs, 122, 109

\bibitem[{{van den Bergh} {et~al.}(2002){van den Bergh}, {Abraham}, {Whyte},
  {Merrifield}, {Eskridge}, {Frogel}, \& {Pogge}}]{vdbergh2002}
{van den Bergh}, S., {Abraham}, R.~G., {Whyte}, L.~F., {Merrifield}, M.~R.,
  {Eskridge}, P.~B., {Frogel}, J.~A., \& {Pogge}, R. 2002, \aj, 123, 2913

\bibitem[{{York} {et~al.}(2000){York}, {Adelman}, {Anderson}, {Anderson},
  {Annis}, {Bahcall}, {Bakken}, {Barkhouser}, {Bastian}, {et~al.}}]{york00}
{York}, D.~G., {Adelman}, J., {Anderson}, J.~E., {Anderson}, S.~F., {Annis},
  J., {Bahcall}, N.~A., {Bakken}, J.~A., {Barkhouser}, R., {Bastian}, S.,
  {et~al.} 2000, AJ, 120, 1579

\end{thebibliography}
\end{document}